\documentclass[twocolumn,amsmath,amssymb]{revtex4}

\usepackage{graphicx}% Include figure files
\usepackage{dcolumn}% Align table columns on decimal point
\usepackage{bm}% bold math

\newcommand\Tm{\langle\mathbf{T}\rangle}

\begin{document}

\title{Comment on ``Localization Transition of Biased Random Walks on Random Networks''}

\author{O. B\'enichou}
\author{R. Voituriez}%\email{voiturie@lptmc.jussieu.fr}
\affiliation{%
Laboratoire de Physique Th\'eorique de la Mati\`ere Condens\'ee,\\
Universit\'e Pierre et Marie Curie, 4 Place Jussieu, 75252 Paris France }

\maketitle

Sood and Grassberger studied in \cite{Sood2007} random walks on random graphs that are biased towards a fixed target point. They put forward  a critical bias strength $b_c$ such that a random walker on an infinite graph eventually reaches the target with probability 1  when $b>b_c$, while a finite fraction of walks drift off to infinity for $b<b_c$. They rely on rigorous results obtained for biased walks on  Galton-Watson (GW) trees \cite{peres} to calculate $b_c$, and give arguments indicating that this result should also hold for random graphs such as Erdos-Renyi (ER) graphs and Molloy-Reed (MR) graphs. To validate their prediction, they show  by numerical simulations that the mean return time (MRT) on a finite ER graph, as a function of the graph size $N$, exhibits a transition around the expected $b_c$.

Here we show that the MRT on a GW tree can actually be computed analytically. This allows us (i)  to show analytically that indeed the MRT displays a transition at $b_c$, (ii) to elucidate the $N$ dependence of the MRT, which contradicts the $\propto N$ scaling expected in \cite{Sood2007} for $b<b_c$.

Let us consider a realization of 
a GW tree of $g$ generations, and denote by $z_n$ the number of nodes of generation $n$ ($0\le n\le g-1$). The number of
nodes at generation zero is taken equal to one ($z_0=1$) and 
 each node has a random number of daughter nodes of mean $k$. As in ref. \cite{Sood2007}, we consider a random walker starting from the root node ($n=0$) and experiencing  
a constant bias, such that the probabilities $p_l^-$ and $p_l^+$ to jump from site $l$ respectively  towards and away from the root are given by $p_l^-=b/{\cal N}_l$ and $p_l^+=b^{-1}/{\cal N}_l$ where ${\cal N}_l$ is a normalization constant. 

The key point of the derivation is that the MRT, here denoted by $\Tm$, is given by the Kac formula \cite{Aldous,Condamin2007,nature2007}:
\begin{equation}
 \Tm=\frac{1}{P_{\rm eq}(0)}, 
\end{equation}
where $P_{\rm eq}(n)$ is the equilibrium distribution at generation $n$ which is easily showed to verify $P_{\rm eq}(n)\propto z_n b^{-2n}$. Normalization then yields straightforwardly
 \begin{equation}
 \Tm=\sum_{n=0}^{g-1}z_n b^{-2n}. 
\end{equation}
We then denote by $\overline{X}$ the average of a quantity $X$ over the realizations of the graph. Using $\overline{z_n}=k^n$ \cite{Feller}, we first obtain the average number of nodes in the GW tree $N=(1-k^g)/(1-k)$, and finally the desired quantity

\begin{equation}\label{Tm}
\overline{\Tm}=\frac{1-(1+(k-1)N)^\epsilon}{1-(k/b^2)}
\end{equation}
with $\epsilon=\ln(k/b^2)/\ln(k)$.

Equation (\ref{Tm}) clearly shows that $\overline{\Tm}$ exhibits a transition when $\epsilon=0$ or equivalently $b=b_c=\sqrt{k}$, in agreement with
the numerical simulations of  \cite{Sood2007}. Furthermore, we  obtain explicitly the  asymptotics of $\overline{\Tm}$ for large $N$:

\begin{equation}\label{Tlim}
\overline{\Tm} \sim \left\{
\begin{array}{ll}
\displaystyle  \frac{(k-1)^\epsilon}{k/b^2 -1}N^\epsilon \ {\rm for}\ \epsilon>0\ (b<b_c) \\
\displaystyle \frac{\ln N}{\ln k}\ {\rm for}\ \epsilon=0\ (b=b_c) \\
\displaystyle\frac{1-(k-1)^\epsilon N^\epsilon}{1-k/b^2 }\ {\rm for}\ \epsilon<0\ (b>b_c)
\end{array}\right.
\end{equation}
Equations (\ref{Tm}) and (\ref{Tlim}) show that in the case of an unbiased walk $(b=1)$, one has  $\overline{\Tm}=N$, but that this scaling does not hold as soon as $b\not=1$.
Note however that this exact scaling could be hard to distinghuish from the numerical simulations of  \cite{Sood2007}.

We thank   the authors of \cite{Sood2007} and acknowledge that they have agreed on the validity of this result and agreed to publication of this Comment.
%This shows in particular that the scaling $\overline{\Tm}\propto N$  does not hold as soon as $b\not=1$.

\end{document}